\documentstyle[aps,preprint,epsfig]{revtex}

\newcommand{\be}{\begin{equation}}
\newcommand{\ee}{\end{equation}}
\newcommand{\bea}{\begin{eqnarray}}
\newcommand{\eea}{\end{eqnarray}}
\newcommand{\bdm}{\begin{displaymath}}
\newcommand{\edm}{\end{displaymath}}

\newcommand{\bi}{\begin{itemize}}
\newcommand{\ei}{\end{itemize}}

\newcommand{\refkl}[1]{(\ref{#1})}

\newcommand{\sub}[1]{_{\rm #1}}
\renewcommand{\sup}[1]{^{\rm #1}}

\newcommand{\msii}{m/s$^2$}

\begin{document}

\tighten
\onecolumn

\title{Microsimulations of Freeway Traffic 
               Including Control Measures}
\author{Martin Treiber and Dirk Helbing}
\address{Institute for Economics and Traffic, TU Dresden,
         Andreas-Schubert-Str. 23, D-01062 Dresden, Germany.\\
{\tt http://www.helbing.org}
}
\maketitle

\begin{abstract}
Using a recently developed microscopic traffic model,
we simulate how speed limits,
on-ramp controls, and
vehicle-based driver-assistance systems influence
freeway traffic. 
We present results for a section of the German Autobahn
A8-East.
Both, a speed limit and an on-ramp control could considerably reduce  
the severeness of the originally observed and simulated  congestion.
Introducing 20\% vehicles equipped with
driver-assistance systems eliminated the  congestion almost completely.
\end{abstract}

\section{\label{sec_intro}Introduction}
For many years, now, the volume of vehicular traffic increases continuously.
Lack of space and money, or ecological considerations often do not allow to
respond to this rising demand by expanding the infrastructure.
Control strategies for vehicular traffic offer the possibility
to increase both, the capacity and the stability of traffic flow without building new
streets. 
The reason why such strategies are promising is the fact that
(i) traffic breakdowns are typically triggered
by perturbations which can be reduced by
suitable control measures,
(ii) traffic flow typically drops by about 5\% to 30\%
\cite{Hall1,Persaud} after a breakdown.
This gives a first estimate for the potential capacity gain.

Testing of new control strategies on real traffic is very
expensive and not always feasible. In contrast, traffic simulations allow
to assess the 
performance of a given control strategy in a short time. Therefore,
simulations are
especially useful in selecting the best strategy
during early stages of implementing new traffic controls.

There are several approaches to model vehicular traffic 
which can be used to simulate traffic
controls (see the overview in Ref.~\cite{Helb-Opus}).

{\it Macroscopic} models make use of the picture of
traffic ``flow'' as a physical flow of some fluid.
They describe the dynamics of ``macroscopic'' 
quantities like the traffic density, traffic flow,
or the locally averaged velocity as a function of space and time.
Therefore, they are suitable to model control measures that influence 
directly these quantities, like on-ramp controls \cite{Handbook}.

In contrast, {\it microscopic} models describe the motion
of each individual vehicle, i.e., they model the driving reactions
(accelerations, braking
decelerations, and lane changes) of each driver as a response to the
surrounding traffic.
They are especially suited to model control measures
that influence selectively individual driver-vehicle units,
such as  driver-assistance systems
or speed limits (a speed limit of, e.g., 100 km/h should not affect trucks).
However, to the knowledge of the authors,
no accepted model or systematic simulation study of 
these effects has been published.

Microscopic models have probably the longest history of all traffic models.
Nowadays, there is an enormous variety,
ranging from low-fidelity single-lane
models for academic purposes (see, e.g., the overview in Ref.~\cite{CA}) to 
very detailled high-fidelity multi-lane
models with the goal of simulating traffic
as realistically as possible (see, e.g., Ref.~\cite{Wiedemann-Schwerdtfeger}).

From a physicist's point of view, the simple models are useful
to investigate generic properties of traffic flow
like stop-and-go waves in its purest form. 
The dynamics of individual driver-vehicle units, however, is not modelled 
realistically.

High-fidelity models can model nearly 
every traffic situation including any conceivable control measure.
Unfortunately, their level of detail comes with many model parameters 
(more than 50 are not uncommon), 
which increases the sensitivity to small parameter changes and
complicates calibration.

\section{\label{sec_mic}The Intelligent-Driver Model (IDM)}

In this paper, we  will use the recently developed 
``intelligent-driver model'' (IDM) \cite{Opus}. 
With respect to complexity, it lies in between the simplistic and
high-fidelity models mentioned above.
For simplicity, we will simulate a single-lane main road and use a
very simple lane-change model for on-ramps. 
Therefore, we do not consider situations where lane changes on the
main road are
important like weaving sections.
In fact, theoretical
investigations backed up with empirical data \cite{Opus}
have shown that for many types of road inhomogeneities
lane-changes on the main road
do not play an important role. 

The seven parameters of the IDM are intuitive and have
plausible values, cf. Table \ref{tab_param}.
We have shown that
this model describes realistically both, the driving behavior
of {\it individual} drivers and the {\it collective}
dynamics of traffic flow like stop-and go waves, or the aforementioned
capacity drop \cite{Opus,Treiber-TGF99}.

The IDM acceleration $\dot{v}$ of each vehicle is 
a continuous function of its own 
velocity $v$, the spatial gap $s$ 
to the leading vehicle, and the
velocity difference (approaching rate)
$\Delta v$ to the front vehicle:
\begin{equation}
\label{IDMv}
\dot{v} = a
         \left[ 1 -\left( \frac{v}{v_0} 
                  \right)^{\delta} 
                  -\left( \frac{s^*(v,\Delta v)}
                                {s} \right)^2
         \right].
\end{equation}
This expression is an interpolation of the tendency to accelerate 
on a free road according to the formula
$a[1-(v/v_0)^{\delta}]$
and the tendency to brake with deceleration
$-a(s^*/s)^2$, 
when the vehicle comes too
close to the vehicle in front. 
The deceleration term
depends on the ratio between the ``desired
minimum gap'' $s^*$ and the actual gap $s$, where the desired gap
\begin{equation}
\label{sstar}
s^*(v, \Delta v) 
    = s_0 
    + s_1 \sqrt{\frac{v}{v_0}}
    + T v
    + \frac{v \Delta v }  {2\sqrt{a b}}
\end{equation}
is dynamically varying with the velocity $v$ and the approaching
rate $\Delta v$.

In general, the IDM parameters 
have different values for each individual vehicle
representing different
driving styles and motorizations (see below).
Many aspects of traffic control, however, can be simulated assuming 
two driver-vehicle classes
(two parameter sets representing, e.g., cars and trucks). 
For simplicity, we will restrict ourselves to these cases, here.

\subsection{\label{sec_param}Model Properties}

The IDM accceleration \refkl{IDMv}
is a continuous function of the
variables $s$, $v$ and $\Delta v$ describing the 
traffic situation seen by the driver.
The driving styles implemented by the IDM  can be seen when considering
the following limiting cases:
\bi
\item
On a {\it nearly empty freeway} corresponding to $s\gg v_0 T$, 
the acceleration is given by $\dot{v}=a[1-(v/v_0)^{\delta}]$.
The driver accelerates to his desired velocity $v_0$
with the  maximum
acceleration given by $a$.
The acceleration coefficient $\delta$ influences the changes of the 
acceleration when
approaching $v_0$.
For $\delta=1$, we have an exponential approach 
with a relaxation time of $\tau=v_0/a$. In the limit $\delta\to\infty$,
the acceleration $\dot{v}=a$ is constant
during the whole acceleration process and drops to zero
when reaching $v_0$.
\item
In {\it dense equilibrium traffic} corresponding to $\dot{v}=0$ and 
$v < v_0/2$,
drivers follow each other with a constant distance 
$s_e(v) \approx s^*(v,0) = s_0 + s_1\sqrt(v/v_0) + v T$.
For the case $s_1=0$ considered here, the distance is equal to a
small contribution $s_0$ denoting the
minimum bumper-to-bumper distance kept in standing traffic plus a
velocity-dependent contribution $v T$ corresponding to a time headway $T$.
\item
When {\it approaching standing obstacles} from an initially large distance,
i.e., for $\Delta v=-v$ and $s\gg v^2/(2 b)$, IDM drivers brake in a way
that the comfortable deceleration $b$ will not be exceeded
in the approaching phase.
\item
Finally, when there is an {\it emergency situation} characterized by
$s < (\Delta v)^2/(2 b)$, the drivers brake hard 
to get their vehicle under
control, again. 
One could also incorporate a maximum physical deceleration (blocking wheels)
of about 9 \msii, but such decelerations were not reached in our simulations.
\ei

In view of applications in traffic control, it is essential that
the road capacity per lane increases strongly with 
{\it decreasing} time headway $T$ (since the theoretical
upper limit of traffic flow for vehicles of zero length is given by $1/T$),
while traffic stability is enhanced with {\it increasing} $T$ 
(as there is more time to react to new situations).
Furthermore, stability increases with growing $a$ 
(because drivers adapt faster to new situations)
and decreasing $b$ (due to their more anticipative braking reactions). 
As one expects intuitively, the  capacity drop mentioned above
is caused mainly by drivers who accelerate too slowly
when leaving congested traffic. In accordance with this picture, the simulated capacity drop
increases with decreasing acceleration $a$.

\begin{figure}
\includegraphics[width=140mm]{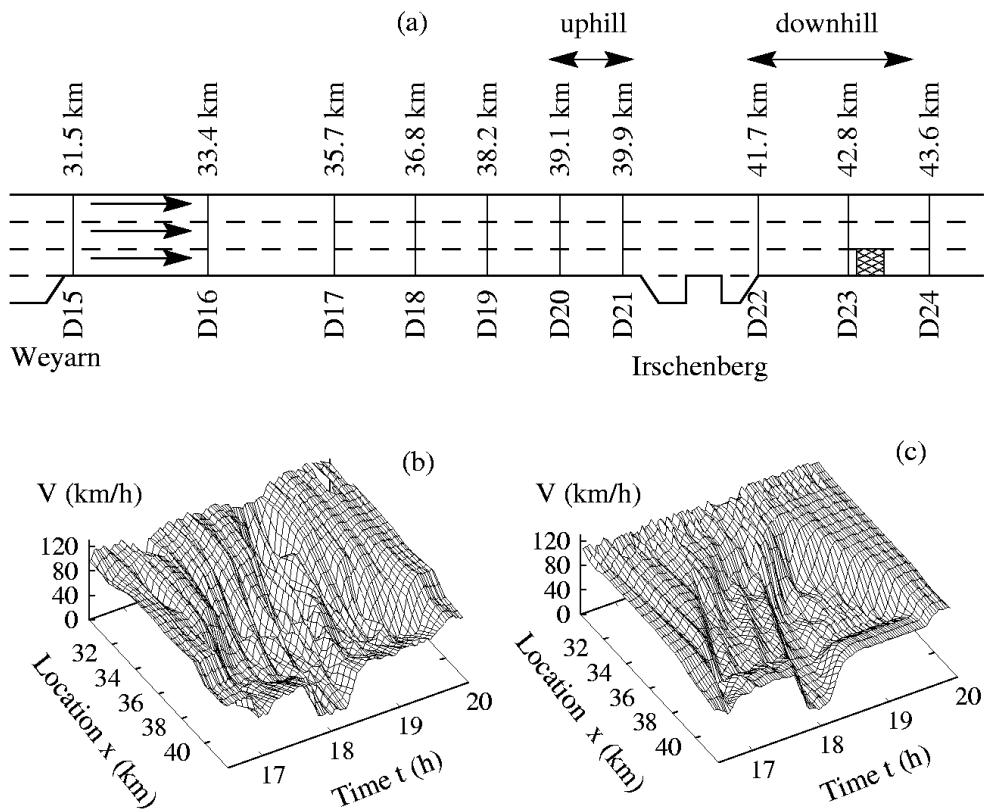}
\caption{\label{fig_A8}
Traffic breakdown at a  section of the A8-East
used as test szenario. (a) Sketch of the section. 
The square
downstream of D 23 indicates a temporary closing of the right lane due
to an incident (see text).
(b) Measured and
(c) simulated locally averaged velocity.}
\end{figure}

\section{\label{sec_control}Traffic Control} 

Traffic controls influence the driving behavior of
some or all vehicles. Some examples are
\bi
\item speed limits, 
\item on-ramp controls (``ramp metering''),
\item dynamic-route guidance systems, and
\item lane-changing restrictions.
\ei
In addition to such externally imposed measures,  a vehicle-based 
automated acceleration control 
is possible for vehicles equipped with driver-assistance systems.

Traffic control by ramp metering and route guidance
is an old and large field ( 
for an overview, see, e.g., \cite{Lapierre-Steierwald,Handbook}), while
comparatively little has been published about modelling of speed limits
\cite{Lapierre-Steierwald,Cremer79,Smulders90,Kuehne-opt,Lenz99}.
In most cases, the effects of the above control measures on the {\it static}
capacity have been investigated.
We are not aware of a systematic simulation
of the {\it dynamic} effects caused by various control measures
using a single micromodel. We will show that dynamic
effects are not only relevant but sometimes produce even
counter-intuitive
results.
In particular, we will show that
speed limits can {\it increase} the dynamic capacity although the
average static capacity is even {\it reduced}.

In the following, we will simulate effects of 
a speed limit, an on-ramp control, and
a driver-assistance system
for a section of the German autobahn A8-East 
containing an uphill section around $x=40$ km.
As further inhomogeneity, there is a small junction at about
$x=41.0$ km. However, since the involved ramp flows were very small,
we assumed
that the junction had no dynamical effect.

We considered the situation
during the evening rush hour
{on} November 2, 1998.
At about 17 h, traffic broke down at the uphill section.
At about 18 h,
an additional  congestion caused by an incident further downstream
propagated
into the simulated section \cite{Opus}.
In the simulation shown in Fig. \ref{fig_A8}(c), we used measured 
lane-averaged one-minute data of velocity and flow as 
upstream and downstream boundary conditions.
Both types of traffic breakdowns were realistically reproduced.
For the purpose of simulating traffic controls, however, we will
exclude the incident-caused jam, here, replacing the
data-based downstream boundary conditions by ``free'' 
boundary conditions \cite{numerics}.
The reason is that congested traffic caused by accidents 
can be hardly eliminated by control measures. (The probability for 
an incident to occur can certainly be influenced,
but this will not be discussed here).

\begin{figure}
\begin{center}
\includegraphics[width=140mm]{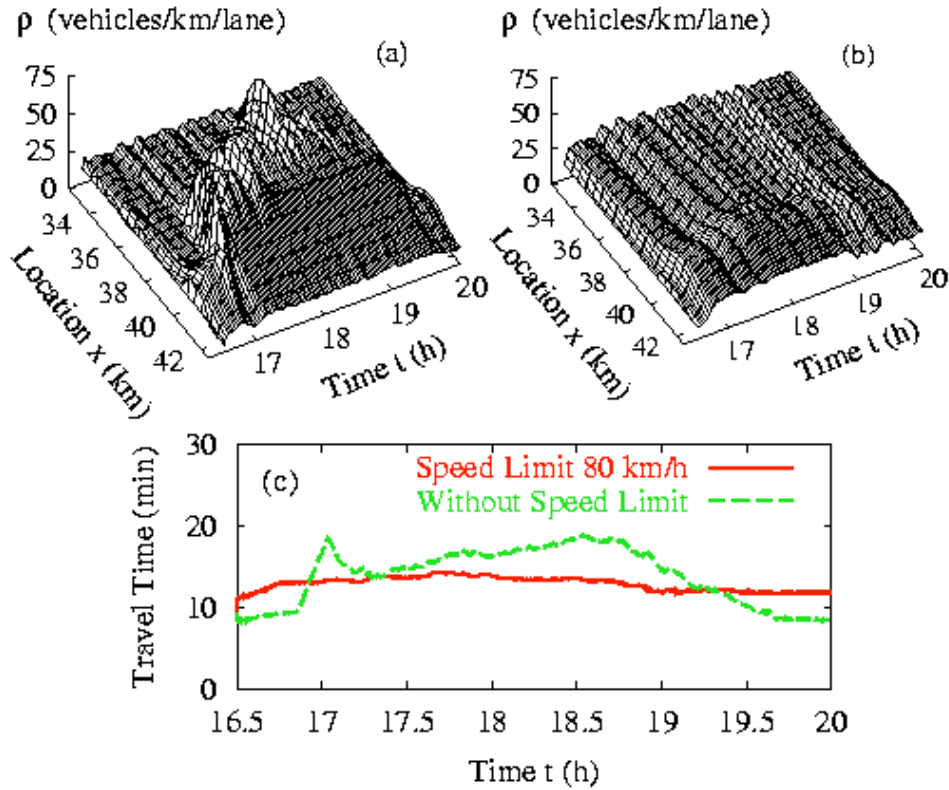}
\end{center}
\caption{\label{speedl}
Simulation of the A8-East (a) without speed limit, (b)
with a speed limit of 80 km/h. (c) Travel times corresponding to the scenarios (a) and (b).
}
\end{figure}

\subsection{\label{sec_speedl}Speed limits}

We have assumed two vehicle classes:
50\% of the drivers had a desired velocity of $v_0=120$ km/h, 
while the other half had  $v_0=160$ km/h outside of the uphill
region. A speed limit reduces the desired velocities to 80 km/h.
Within the uphill region, both driver-vehicle classes are 
forced to drive at a maximum of
60 km/h (for example, due to overtaking trucks
that are not
considered explicitely here; their 
realistic inclusion requires multi-lane simulations.)

Figure \ref{speedl} shows spatio-temporal plots of the 
locally averaged traffic density for scenarios
with and without the speed limit. (For the specific averaging formula, see Ref.~\cite{Opus}.)
The simulations show the following:
\bi
\item During the rush hour (17 h $\le t\le 19$ h), the overall effect of the
  speed limit is positive.
The increased travel times in regions without congestion
are overcompensated by the saved time due to the avoided
breakdown.
\item
For lighter traffic ($t<17$~h or $t>19$:30~h), 
however, the effect of the speed limit clearly is negative.
This problem can be circumvented by traffic-dependent,
variable speed limits.
\ei

\begin{figure}
\includegraphics[width=140mm]
  {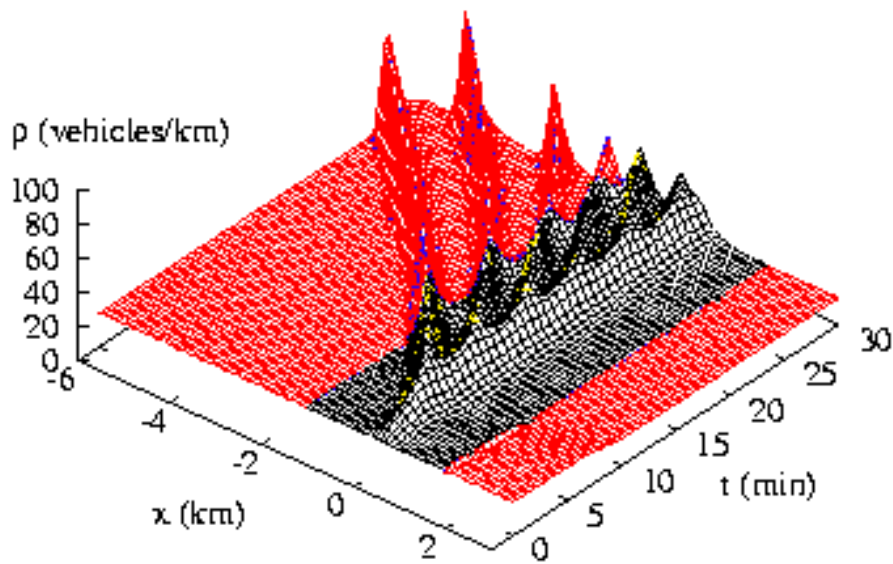}
\caption{\label{fig_micmac}
Simulation of an on-ramp with the micro-macro link
(from \protect\cite{Hennecke-TGF99}). The on-ramp region (dark)
is computed with a macroscopic version of the IDM.
}
\end{figure}

\subsection{\label{sec_rmp}On-Ramp Control}
%
Traffic inflow on
on-ramps can be controlled, e.g.,  by traffic lights (``ramp metering'').
Many on-ramp flow-control strategies have been proposed in the last 30 years
and applied mostly to macroscopic simulations
\cite{Handbook}.
Clearly, lane-changing processes play an important role in simulating
on-ramp flow-control \cite{Lapierre-Steierwald}. The role of the
longitudinal dynamics (acceleration and braking), however, 
remained less clear.

Here, we will show that on-ramp control can be beneficial even when
considering essentially longitudinal effects.
To this purpose, we neglect lane changes on the main road and
apply following minimal lane-changing scheme
to implement an on-ramp of  length $L\sub{rmp}$ 
from which a traffic flow $Q\sub{rmp}$ merges  to a
single main lane:
\bi
\item
Try to add ramp vehicles to the main flow at the times 
when the temporal ramp-flow integral
reaches integer values.
\item Search the largest gap on the main lane in the section of length $L_{\rm rmp}$ adjacent to the
  acceleration lane.
\item If the resulting 
gaps to the respective front and rear vehicles exceed a certain minimal value, 
place the new vehicle in the middle with an initial velocity given by
the average of 
the front
and rear vehicles.
\item Otherwise, add the vehicle to the queue of waiting vehicles.
(This case did not occur in the simulations. All queues were caused by the
flow control to be described below).
\ei
Another possibility (apart from a multi-lane model)
is opened by the recently formulated micro-macro link
\cite{Hennecke-TGF99} allowing to 
simulate the ramp section macroscopically with a source term in
the continuity equation \cite{sync-Letter}, while the
remaining stretch is simulated microscopically, see Fig.~\ref{fig_micmac}.

We restrict the sum of the traffic flow $Q\sub{main}$ upstream of the
on-ramp and the flow $Q\sub{rmp}$ arriving at the on-ramp by
following simple control scheme:
\be
Q\sup{max}\sub{rmp} = \max(0, Q\sub{c}-Q\sub{main}),
\ee
where $Q\sub{main}$ is the traffic flow on the main road upstream of the
on-ramp.
If, at some time $t_{\rm c}$,  the flow $Q\sub{rmp}$ arriving at the on-ramp
exceeds this limit,
a queue of 
\be
n\sub{wait}(t)=\int^t_{t_c} \! dt \, (Q\sub{rmp}-Q\sup{max}\sub{rmp})
\ee
waiting vehicles will form.
The {\it actual} 
inflow $Q^c\sub{rmp}$ of the controlled on-ramp is given by
\be
Q^c\sub{rmp} = \left\{ \begin{array}{ll}
  Q\sup{max}\sub{rmp} \ \ & \mbox{if } n\sub{wait}(t)\ge 1,\\
  \min(Q\sub{rmp}, Q\sup{max}\sub{rmp}) \ \ & \mbox{otherwise.}
\end{array}\right.
\ee
Notice that the on-ramp can be completely closed according to this
scheme.
More elaborate and realistic control schemes 
have been proposed in the literature \cite{Handbook}. 
The simulation results, however, are 
robust with respect to the choice of a certain  control scheme,
so we chose the conceivably most simple one.


We simulated single-lane single-class traffic
and modelled the uphill gradient by reducing the
desired velocity from 120 km/h to 60 km/h.

\begin{figure}
\begin{center}
\includegraphics[width=70mm]
   {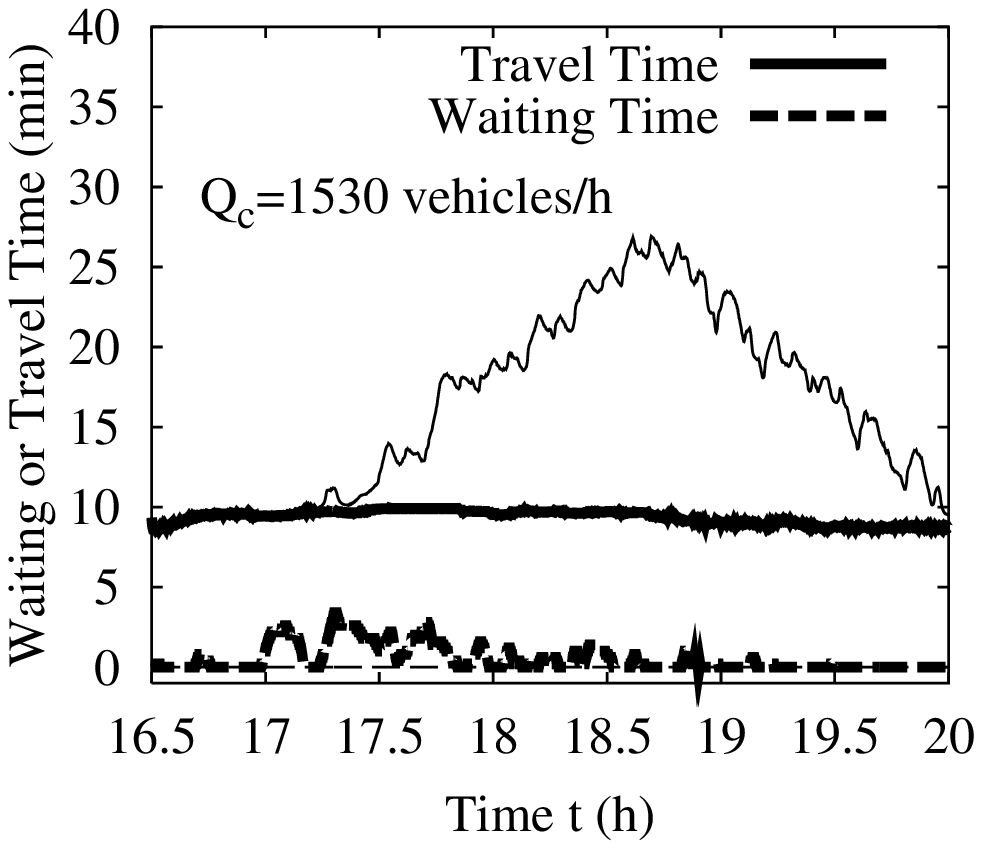} 
\hspace*{-2mm}
\includegraphics[width=70mm]
   {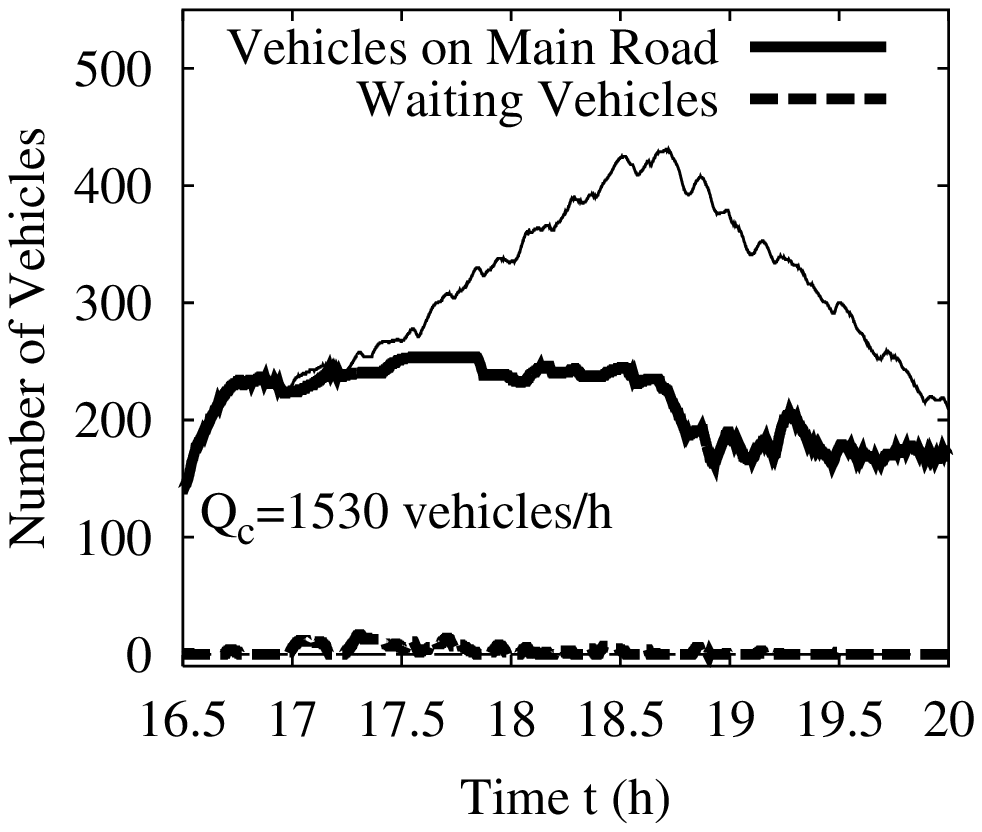} 
\end{center}
\caption{\label{ctrlflow_Opt}
(a) Travel times on the main road (solid lines) and waiting times on the ramp (dashed) for
$Q\sub{rmp}=300$ vehicles/h.
(b) Number of vehicles on the considered section of the main road and number
of waiting on-ramp vehicles. All results are plotted
for optimized flow control (thick lines),
and without flow control (thin lines).
}
\end{figure}

\begin{figure}
\begin{center}
\includegraphics[width=70mm]
   {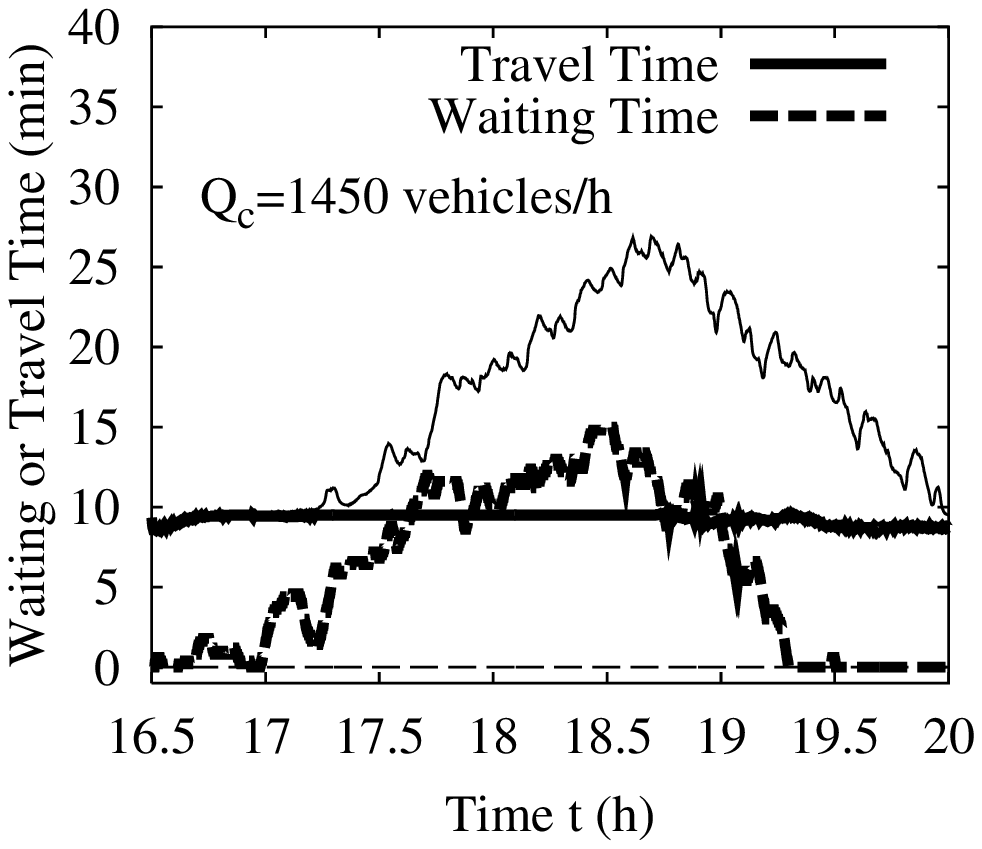} 
\hspace*{-2mm}
\includegraphics[width=70mm]
   {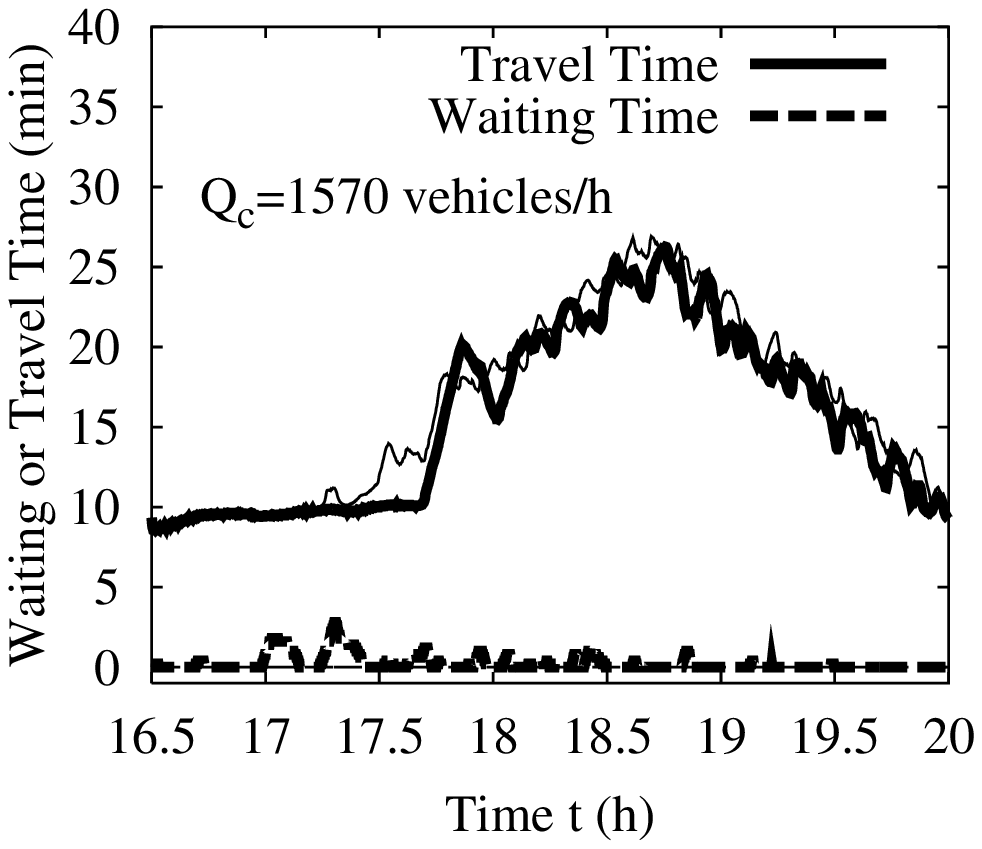} \\[0mm]
\end{center}
\caption{\label{ctrlflow_LowHigh}
Travel and waiting times as in Fig. \protect\ref{ctrlflow_Opt},
but for non-optimal values of the cut-off flow $Q\sub{c}$.
}
\end{figure}

Figure \ref{ctrlflow_Opt}
shows the travel time  through the
simulated section on the main road, and the waiting time at the
on-ramp for an assumed ramp flow $Q\sub{rmp}$ of 300 vehicles per
hour and main lane for the optimal choice of $Q\sub{c}$.
Notice that in the optimized control scenario,  the
waiting times of the on-ramp vehicles are {\it overcompensated}
 by the decreased
travel time for the main road stretch
(with the exception of a short time interval after 17~h).
Since also vehicles coming from the on-ramp drive on the main section 
afterwards, even they will generally profit from 
the flow control in this example.

Figure \ref{ctrlflow_LowHigh} shows the travel and waiting times
 for non-optimal values of $Q\sub{c}$. If
$Q\sub{c}$ is 5\% below the optimal value, the 
maximum waiting time is increased and becomes comparable to the saved 
travel time (Fig. \ref{ctrlflow_LowHigh} left). 
If one increases $Q\sub{c}$ above the optimal value, the 
control becomes less and less effective (Fig. \ref{ctrlflow_LowHigh} right).

A suitable measure for assessing the quality of a traffic control
is given by
the sum of the waiting and travel times of all vehicles during the
whole rush hour 
\cite{Handbook},
\be
\label{Ttot}
T\sub{t} = \int^{t_2}_{t_1} \! dt \, [
  n\sub{wait}(t) + n\sub{main}(t) ] ,
\ee
where $n\sub{main}$ is the number of all vehicles in the considered section
$x_1\le x\le x_2$.
The spatio-temporal region 
$[t_1,t_2]\times [x_1,x_2]$ must be selected such that there is
free traffic at all boundaries.
Notice that the vehicle {\it numbers} and not the travel and waiting
times  enter into the quality measure $T\sub{t}$.
Furthermore, if $Q\sub{main}(t)\le Q\sub{c}$ is satisfied for 
$t_1\le x\le t_2$,  $T\sub{t}$ does not depend on the on-ramp
flow while the waiting times do.
The right part of Fig. \ref{ctrlflow_Opt} illustrates the 
changes of the quality measure $T\sub{t}$  induced by the control, 
see the difference between the areas enclosed by the two solid curves
(saved total travel time for the main traffic) and
by the dashed curve and the $x$-axis
(total waiting time of the on-ramp traffic if the control is active).

\begin{figure}



\hspace*{-5mm}
\includegraphics[width=140mm]{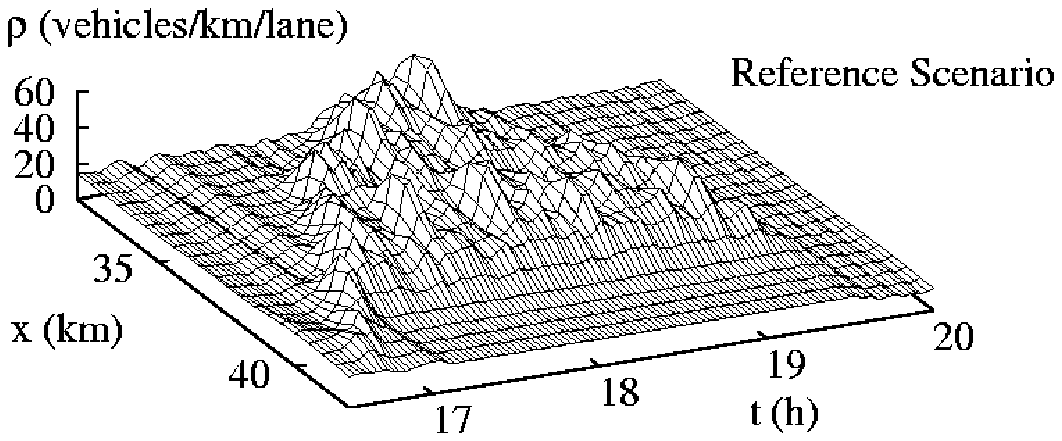} \\[-6mm]
\hspace*{-5mm}
\includegraphics[width=140mm]{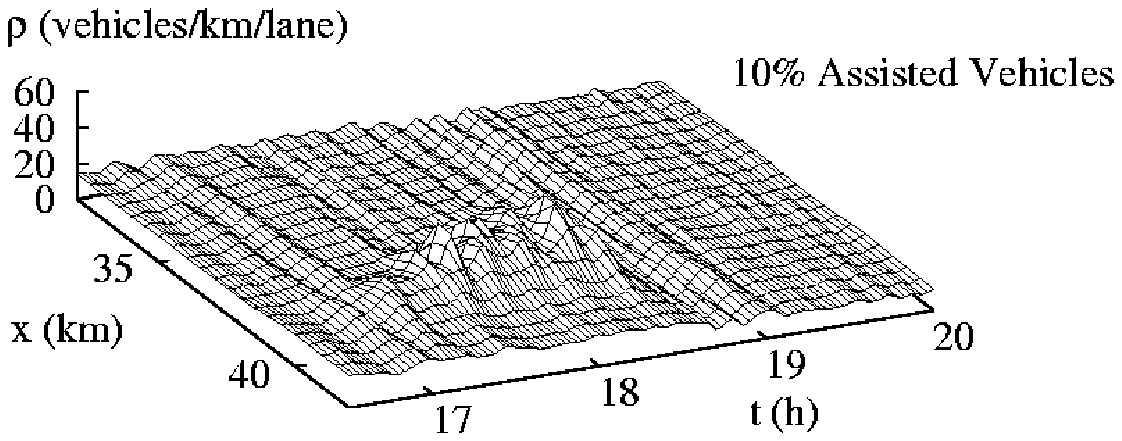} \\[-6mm]
\hspace*{-5mm}
\includegraphics[width=140mm]{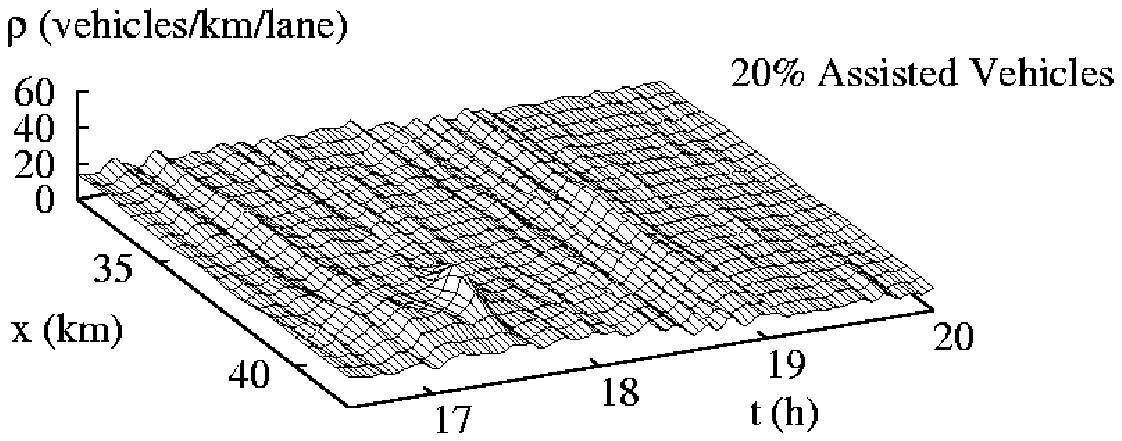} 

\caption{\label{ctrl_FAS}{\small
Spatio-temporal dynamics of the traffic density 
for different percentages of equipped vehicles.
The acceleration of the equipped vehicles has been increased from
$a=1$ \msii\ to $a=2$ \msii. 
The time headway has been decreased from
$T=1.6$ s to $T=0.8$ s.
}}
\end{figure}

\begin{figure}
\begin{center}
\includegraphics[width=140mm]
   {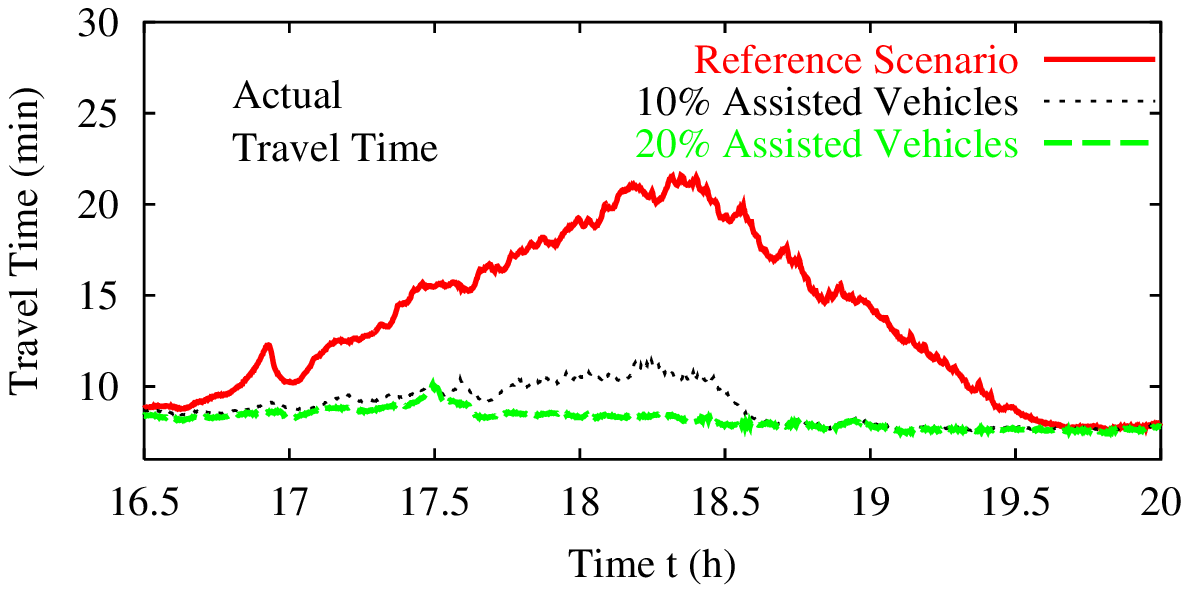}\\[0mm]
\includegraphics[width=140mm]
   {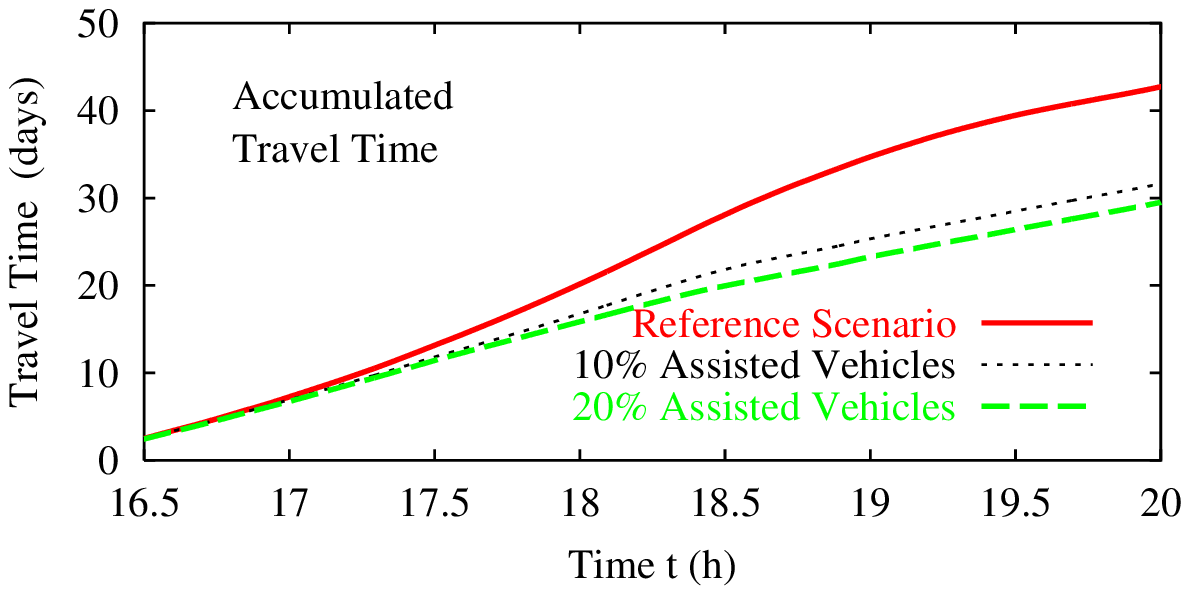}
\end{center}
\caption{\label{FAS_travtime}
(a) Travel time of individual vehicles. 
(b) Total travel time of all vehicles for different percentages
of equipped vehicles.
}
\end{figure}

\subsection{\label{sec_FAS}Vehicle-Based Methods} 

The research presented in this section was performed in tight collaboration
with the Volkswagen AG.

Presently, vehicle-based driver-assistance systems with
adaptive acceleration control become available.
In such systems, a detector determines the distance and velocity difference
to the front vehicle and automatically accelerates or decelerates the vehicle
according to the traffic situation.
Obviously, micromodels like the IDM can be used to implement the 
control algorithm of such a system.

It may be argued that human and automated driving behaviours are
fundamentally different. In fact, the assumptions of 
micromodels like the IDM: reaction only to the immediatel predecessor
in combination with  a 
negligible reaction time, fit perfectly to automated controls but less
to human drivers. Human drivers have a non-negligible reaction time
(of the order of 1 s) leading to more instability.
This is offset by the more elaborate human driving strategy
including direct reaction to decelerations (i.e., to braking lights) and
anticipation of the traffic situations several vehicles in front
of them. Nevertheless, the IDM can be calibrated to
actual traffic data (cf. Fig. \ref{fig_A8}) suggesting that these two
effects not contained in the IDM essentially cancel each other. This
justifies the use of the same model for {\it both} human and automated
longitudinal control.

Here, we show that, in principle,
such control systems can improve the
capacity and quality of traffic flow.

In the simulations, we model both equipped and not equipped vehicles with the
IDM, using different values for $a$ and $T$. 
From the discussion of the IDM in Section \ref{sec_param}
one might expect
that increasing $a$ for the equipped vehicles (within the range allowed by the
motorization) increases stability of the {\it overall} traffic, 
while decreasing $T$
increases its capacity.

Figures~\ref{ctrl_FAS} and \ref{FAS_travtime}
show that this is true 
even for only 10\% equipped vehicles.
The total additional time spent in the traffic jam was decreased by
more than 80\% with respect to the original situation.
For a percentage of 20\% equipped vehicles, the traffic breakdown disappeared
almost completely.

\section{Outlook and Discussion}

In this paper, we have simulated the effects of speed
limits, on-ramp controls, and driver-assistance systems
using the microscopic intelligent-driver model (IDM).
The IDM is suitable for simulating traffic controls
because it is relatively simple,
contains only a few and intuitive model parameters,
and reproduces all relevant properties of real traffic.
Moreover, it is
numerically efficient and robust.

In all simulations, the model parameters have been calibrated to
the historic traffic data of the respective reference scenarios. 
The presented effects of traffic control, however,
are of a qualitative nature. They
need to be calibrated before appying them in real control
systems.

While our simulations of an on-ramp control support the findings
obtained with macroscopic models 
\cite{Handbook,Lapierre-Steierwald},
we show that speed limits and driver-assistance systems have
a similar beneficial potential. 
We emphasize that all three control 
measures have been simulated with the same
model. Thus, it is possible to exploit the synergy effects that might
result from a combination of these measures. 

Both uphill gradients and speed limits reduce the velocity, so,
at first sight, it is puzzling why speed limits can improve the quality of
traffic flow while obviously uphill regions deteriorate it.
To understand this,
note that 
the desired velocity $v_0$ corresponds to the lowest value of
\bi
\item the maximum velocity allowed by the motorization,
\item the imposed speed limits, and
 (possibly with a ``disobedience factor''),
\item the velocity actually ``desired'' by the driver.
\ei
Therefore, 
speed limits act selectively on the  {\it faster} vehicles, 
while uphill gradients reduce predominantly 
the desired velocity of the {\it slower}
vehicles. As a consequence, speed limits reduce velocity differences, 
thereby stabilizing traffic, while uphill gradients increase them.  
Since global speed limits always raise the travel time
in off-peak hours when free 
traffic is unconditionally stable, traffic-dependent speed limits 
are an optimal solution.

In contrast to speed limits, on-ramp controls are
not common in Europe. 
We showed that, nevertheless, they
can help to avoid or delay traffic breakdowns.
Moreover, they can be advantageous even to the drivers who
have to wait at the on-ramps because their total travel times are decreased
as well.

The effects of on-ramp controls are qualitatively comparable to
those of traffic-dependent, dynamic route guidance systems.
While the former truncate 
flow peaks on a short time scale of the order of minutes,
dynamic-route guidance systems truncate flow peaks on a time scale of hours.

In summary, our simulations support the conclusion that control
measures which homogenize traffic flow are generally suitable for the
reduction of congestion and travel times.
While speed limits decrease speed differences,
on-ramp controls smooth out flow peaks.

On the long run, however, vehicle-based systems have certainly 
the highest potential for increasing the traffic capacity of a given infrastructure. 
Driver-assistance systems can improve and speed up the reaction
to the behavior of the respective vehicle in front.
This allows stable traffic at significantly reduced 
time headways. In our test scenario, 20\% equipped vehicles
nearly eliminated all breakdowns.
However, some legislatory and sensor-related problems remain to be solved.

Presently, we are extending our
simulations to multi-lane traffic for the simulation
of further control measures like lane-changing restrictions 
or overtaking bans for all or certain classes of vehicles (e.g., trucks).
Moreover, we are exploring the potentials of communicating vehicles
with respect to their optimal self-organization based on
the paradigms of decentralized control and 
collective intelligence \cite{Helb-optimality}.



\begin{thebibliography}{10}

\bibitem{Hall1}
F.~L. Hall and K. Agyemang-Duah, 
\tit{Freeway capacity drop and the definition of capacity},
Transportation Research Record {\bf 1320},  91
   (1991).

\bibitem{Persaud}
B. Persaud, S. Yagar, and R. Brownlee, 
\tit{Exploration of the breakdown phenomenon in freeway traffic},
Transportation Research Record {\bf
  1634},  64  (1998).

\bibitem{Helb-Opus}
D. Helbing, 
Traffic and Related Self-Driven Many-Particle Systems,
e-print http://arXiv.org/abs/cond-mat/0012229, 
to appear in Reviews of Modern Physics.

\bibitem{Handbook} M. Papageorgiou, in 
{\em Handbook of Transportation Science}, 
edited by R.~W. Hall 
(Kluwer Academic Publishers, London, 1999).

\bibitem{CA} D. Chowdhury, L. Santen, and A. Schadschneider, 
\tit{Statistical physics of vehicular traffic and some related systems},
{Physics Reports} {\bf 329}, {199} (2000). 

\bibitem{Wiedemann-Schwerdtfeger}
T.~S. R.~Wiedemann, 
\tit{Makroskopisches Simulationsmodel f\"ur 
                Schnellstra{\ss}ennetze mit Ber"ucksichtigung von 
                Einzelfahrzeugen (DYNEMO)},
{\em Stra{\ss}enbau und
  Stra{\ss}enverkehrstechnik}, Heft 500 (Bundesministerium f\"ur Verkehr, Abt.
  Stra{\ss}enbau, Bonn-Bad Godesberg, 1987).

\bibitem{Opus}
M. Treiber, A. Hennecke, and D. Helbing, 
\tit{Congested Traffic States in Empirical Observations
               and Microscopic Simulations},
Physical Review E {\bf 62},  1805
  (2000).

\bibitem{Treiber-TGF99}
M. Treiber, A. Hennecke, and D. Helbing,  
\tit{Microscopic Simulation of Congested Traffic},
in {\em Traffic and Granular Flow
  '99}, edited by D. Helbing, H.~J. Herrmann, M. Schreckenberg, and D.~E. Wolf
  (Springer, Berlin, 2000), pp.\ 365--376.

\bibitem{Lapierre-Steierwald}
R. Lapierre and G. Steierwald, 
{ Verkehrsleittechnik f\"ur den Stra{\ss}enverkehr, Band I:
Grundlagen und Technologien der Verkehrsleittechnik}
(Springer, Berlin, 1987).

\bibitem{Cremer79} 
M. Cremer,
{ Der Verkehrsflu{\ss} auf Schnellstra{\ss}en}
(Springer, Berlin, 1979). 

\bibitem{Smulders90}
S. Smulders, S., 1990,
\tit{Control of freeway traffic flow by variable speed signs},
Transpn. Res. {\bf B 24}, 111 (1990). 


\bibitem{Kuehne-opt}
R. D. K{\"u}hne, 
\tit{Freeway control using a dynamic traffic flow model and vehicle
                reidentification techniques},
Transportation Research Record {\bf 1320},
251 (1991).

\bibitem{Lenz99}
H. Lenz,
{Entwicklung nichtlinearer, diskreter Regler zum Abbau 
von Verkehrsflu{\ss}inhomogenit\"aten mithilfe makroskopischer Verkehrsmodelle}
(Shaker Verlag, Aachen, 1999).

\bibitem{numerics}
D. Helbing and M. Treiber, 
\tit{Numerical simulation of macroscopic traffic equations},
Computing in Science and Engineering (CiSE) {\bf 5},
   89  (1999).


\bibitem{Hennecke-TGF99}
A. Hennecke, M. Treiber, and D. Helbing,  
\tit{Macroscopic Simulations of Open Systems and Micro-Macro Link},
in {\em Traffic and Granular Flow
  '99}, edited by D. Helbing, H.~J. Herrmann, M. Schreckenberg, and D.~E. Wolf
  (Springer, Berlin, 2000), pp.\ 383--388.

\bibitem{sync-Letter}
D. Helbing and M. Treiber, 
\tit{Gas-Kinetic-Based Traffic Model Explaining Observed
                Hysteretic Phase Transition},
Physical Review Letters {\bf 81},  3042  (1998).

\bibitem{Helb-optimality}
D. Helbing and T. Vicsek, 
\tit{Optimal self-organization},
New Journal of Physics {\bf 1}, 13.1 (1999),
see http://www.njp.org.
\end{thebibliography}
 

{\bf Acknowledgments:}
The authors are grateful for financial support by
the DFG (grant no.~He 2789/2-1).
The research of Section \ref{sec_FAS} 
was initiated and funded by the strategic 
research of the Volkswagen AG, Wolfsburg, and the results were 
produced in tight
collaboration.

\newcommand{\tit}[1]{}     
\renewcommand{\tit}[1]{#1}  



\vspace{0mm}

\newcommand{\entry}[2]{\parbox{45mm}{#1} &
                      \parbox{25mm}{#2} }

\begin{table}

\begin{tabular}{l|l}
\entry{Parameter} {Typical value}  \\[3mm] \hline 
\entry{}{}{} \\[-2mm]
\entry{Desired velocity $v_0$}
     {120 km/h}
     \\[0mm]
\entry{Safe time headway $T$}
     {1.5 s}
     \\[0mm]
\entry{Maximum acceleration $a$}
     {1 m/s$^2$}
     \\[0mm]
\entry{Comfortable deceleration $b$}
     {2 m/s$^2$}
     \\[0mm]
\entry{Minimum distance $s_0$}
     {2 m}
     \\[0mm]
\entry{Jam distance $s_1$}
     {0 m}
     \\[0mm]
\entry{Acceleration exponent $\delta$}
     {4}
\end{tabular}
\vspace*{2mm}
\caption{\label{tab_param} Parameters of the IDM
used as reference values.
}
\end{table}

\end{document}